# Moiré potential, lattice corrugation, and band gap spatial variation in a twist-free MoS$_2$/MoTe$_2$ heterobilayer

W. T. Geng[a*], V. Wang[b], Y. C. Liu[b,a], T. Ohno[a], J. Nara[a†]

[a] *National Institute for Materials Science, Tsukuba 305-0044, Japan.*
[b] *Department of Applied Physics, Xi'an University of Technology, Xi'an 710054, China*

**Abstract**

To have a fully *ab initio* description of the Moiré pattern in a transition metal dichalcogenide heterobilayer, we have carried out density functional theory calculations, taking accounts of both atomic registry in and the lattice corrugation out of the monolayers, on a MoTe$_2$(9×9)/MoS$_2$(10×10) system which has a moderate size of superlattice larger than an exciton yet not large enough to justify a continuum model treatment. We find that the local potential in the midplane of the bilayer displays a conspicuous Moiré pattern. It further leads us to reveal that the variation of the average local potential near Mo atoms in both MoTe$_2$ and MoS$_2$ layers make intralayer Moiré potentials. They are the result of mutual modulation and correlate directly with the spatial variation of the valence band maximum and conduction band minimum. The interlayer Moiré potential, defined as the difference between the two intralayer Moiré potentials, has a depth of 0.11 eV and changes roughly in proportion to the band gap variation in the Moiré cell, which has an amplitude of 0.04 eV. We find the lattice corrugation is significant in both MoTe$_2$ (0.30Å) and MoS$_2$ (0.77Å) layers, yet its effect on the electronic properties is marginal. The wrinkling of the MoTe$_2$/MoS$_2$ bilayer enhances the spatial variation of the local band gap by 5 meV, while its influence on the global band gap is within 1 meV. A simple intralayer band-coupling model is proposed to understand the correlation of Moiré potential and spatial variation of the band gap.

[*] geng.wentong@nims.go.jp
[†] nara.jun@nims.go.jp

## I. INTRODUCTION

The development of heterostructure in semiconductor electronics has often been impeded by lattice mismatch at the interfaces which usually renders dangling chemical bonds, and hence the instability and abrupt variation of the electronic states at the interface. The weak van der Waals (vdW) interactions bonding layered materials together can enable us to avoid this challenge and indeed the vdW epitaxy technique has been introduced in growth of NbSe$_2$ monolayer on a cleaved face of MoS$_2$ in a pioneering work by Koma *et al*. over three decades ago [1]. However, only after the breakthrough in isolation of graphene from graphite [2] has the vdW heterostructures become a perfect two-dimension (2D) platform allowing exploration of novel materials properties [3]. The simplest and also most intensively studied 2D vdW heterostructures are a stacking of two different monolayers, i.e., heterobilayers, in which the Moiré pattern arises due to either lattice difference or rotational misalignment, or both. In contrast to the lattice constants which are intrinsic property of a material and usually can only be slightly altered by stresses, the rotational freedom in the configuration of a bilayer has proved, firstly in graphene systems [4], and later in other heterostructures [5] to be a rich source in generating new quantum phenomena. In this regard, Moiré-pattern-based electronics could be more specifically described as twist-angle-based electronics, or *Twistronics*, a term corned by Carr *et al.* referring to the manipulation of the electronic properties of two-dimensional layered structures via twist angle [6].

The astonishing discovery of superconducting and insulating states in magic-angle twisted graphene [7-8] has not only stimulated intensive interest in understanding how van der Waals forces can modulate the electronic interactions in such a pristine material [9], but also inspired tremendous efforts in exploration of Moiré excitons in structurally similar but chemically a little more complicated bilayers of semiconducting transition-metal dichalcogenides (TMDCs) [10-13]. First-principles density functional theory (DFT) calculations on vdW TMDC heterobilayers have been performed for small superlattices [14-15]. It was found that the electronic structures of constituent monolayer of TMDC are well retained in their respective free-standing monolayers as a consequence of weak interlayer vdW bonding and the calculated band offsets are in good agreement with experiment [15]. We note that the superlattices in these bilayer systems are rather small (1-2 nm) and can hardly host Moiré excitons, which, estimated by solving the Bethe-Salpeter equation based on massive Diract model [16] is about one nm in MoS$_2$ monolayer. Limited by the computational cost, approximations have to be made in dealing with much larger Moiré cells. For twist-free MoS$_2$(24 × 24 )/MoSe$_2$(23 × 23 ), which has a Moiré cell size of 8.7 nm to minimize lattice mismatch, Kang et al. [17] employed the linear scaling three-dimensional fragment method; for similar twist-free systems MoS$_2$/WSe$_2$ and MoSe$_2$/WSe$_2$, DFT calculations were only performed separately for the three regions with three-fold rotational-symmetry, on the smallest 1:1 supercell with an average lattice constant of the two constituent monolayers [18-19]. Since strain has strong influence on the electronic structures of TMDC [20], full DFT treatment is still desirable in order to get more precise details of the Moiré potential which determines the lattice corrugation of the bilayer, the spatial variation of the band gap, and many other quantum phenomena arising from interlayer vdW interactions, especially for a Moiré cell not large enough to fully justify a continuum model which ignore local atomic arrangements [4].

It is noteworthy that although the spatial variation of the potential within a Moiré cell has been known for decades [21], the term *Moiré periodic potential* appeared only very recently in a report of strong evidence of its presence in twisted graphene bilayer [22], and its effects, rather than itself, has been the focus of some immediate follow-up studies on various material systems [23-24]. Here, *variation* of a potential defined in real space means the difference of this potential at equivalent points in each original unit cell of an individual monolayer, rather than the change of a potential within each unit cell. That is, the variable of a Moiré potential has but one element in each original unit cell. In a surface-absorbate system, such a variation was naturally defined in reference to the lattice of the solid surface [25]. In the case of a bilayer, however, the definition of Moiré potential should depend on which parts of the electronic structure, or, which energy bands, are of interest.

To our knowledge, MacDonald's group are the first to evaluate the Moiré potential of vdW bilayers, making use of information obtained from *ab initio* calculations on high-symmetry local crystalline structures without mismatch and twist to derive effective Hamiltonians that are able to efficiently describe the effect of the moiré pattern on electronic properties of the long-period superlattices [26]. The most important inference of its two-band model approximation is that the Moiré potential can be defined by the spatial variation of the band gap as a function of the relative displacement ***d*** (between two monolayers bearing similar lattice constants but misoriented slightly) in a 2D Moiré superlattice:

$$\Delta(d) \equiv E_g(\boldsymbol{d}) - \langle E_g \rangle \approx \sum_{j=1}^{6} V_j \exp(i\boldsymbol{G}_j \cdot \boldsymbol{d}) \qquad (1)$$

where $E_g$ is the intralayer band gap of MoS$_2$, $\langle E_g \rangle$ is its average over ***d*** in the Moiré cell, and $\boldsymbol{G}_j$ are the first-shell reciprocal lattice vectors of the Moiré cell [27]. Note that ***G*** determines the allowed values for ***d***. This model performs very well on twisted homobilayers [28] or heterobilayers with negligible lattice mismatch [27, 29]. If the constituent monolayers differ much in lattice constants, such as in the TMDC bilayer having different chalcogen elements, this effective model will not be applicable because strain has strong influence on the electronic structures of graphene [30], TMDC [20] and presumably many other kinds of monolayers. It is worth noting that the Moiré potential defined this way is only available after the band gap of the Moiré cell has been determined.

Apparently, fully *ab initio* description of the Moiré pattern in vdW heterobilayers, taking accounts of both atomic registry in and the lattice corrugation out of the atomic planes, is highly desirable. Here, we attempt to study the Moiré pattern dependence of structural and electronic properties of a vdW heterobilayer composed of two prototype TMDC monolayers using DFT calculations. To keep the computational cost affordable, we choose the MoTe$_2$/MoS$_2$ bilayer as an example in view that MoTe$_2$ ($a_1$=3.528 Å) and MoS$_2$ ($a_2$=3.164 Å) monolayers [31] have a remarkable difference ($a_1/a_2$=1.115) in their lattice constants and thus the superlattice of the bilayer has a moderate size. This heterobilayer has been studied by Kou et al. to investigate the effect of strain and spontaneous polarization on the band gap for a series of TMDC hetero-multilayers. [32]

## II. COMPUTATIONAL DETAILS

The misfit in a twist-free MoTe$_2$(n×n)/MoS$_2$[(n+1)×(n+1)] bilayer is $\delta a = \frac{3.528 \times n}{3.164 \times (n+1)} - 1$. Thus, a MoTe$_2$(9×9)/MoS$_2$(10×10) stacking has the least supercell lattice mismatch $\delta a$=0.35%. When grown epitaxially on graphene, TMDC monolayers bear lattice constants very close to those of isolated monolayers, albeit with various twist angles [33]. Since the vdW interactions are expected to be stronger in a TMDC/TMDC bilayer than in a TMDC/graphene bilayer, we are curious whether lattice misfit in the former case will produce strain in the individual monolayers. To obtain the energy-strain relationship, we have also investigated the (7×7)/(8×8), (8×8)/(9×9), and (10×10)/(11×11) stackings, which have a *nominal* (before 2D optimization) supercell lattice mismatch $\delta a$ of -2.59%, -1.04%, and 1.21% respectively. We use the word *nominal* to stress that in a bilayer system, the lattice strain after 2D optimization is not well defined because there is no sound reference system for each monolayer whose lattice constants are different in free-standing and bulk form. Furthermore, to assess the effect of extraordinarily large lattice mismatch, we have also calculated the (1×1)/(1×1) stacking with an average lattice constant (3.346 Å) of the two monolayers. In each supercell, the length of *c* axis was set to 30 Å, yielding a vacuum region of around 23 Å, large enough to minimize the interactions between mirror bilayers. The length in *a*, on the other hand, has been optimized.

The first-principles density functional theory (DFT) computation was performed using Vienna Ab initio Simulation Package. [34] The electron-ion interaction was described using projector augmented wave (PAW) method. [35] The exchange correlation between electrons was treated both with generalized gradient approximation (GGA) in the Perdew-Burke-Ernzerhof (PBE) form [36]. The non-bonding vdW interaction was incorporated by employing a semi-empirical correction scheme of Grimme's DFT-D3 method [37]. We used an energy cutoff of 400 eV for the plane wave basis set for all systems, both monolayers and bilayers, to ensure equal footing. The Brillouin-zone integration was performed within Monkhorst-Pack scheme using *k* meshes of (12×12×1) for (1×1) monolayer unit cells and (1×1)/(1×1) bilayer supercells, (2×2×1) for $MoTe_2$(7×7)/$MoS_2$(8×8) and $MoTe_2$(8×8)/$MoS_2$(9×9), and (1×1×1) ($\Gamma$ point only) for $MoTe_2$(9×9)/$MoS_2$(10×10) and $MoTe_2$(10×10)/$MoS_2$(11×11) stackings. A finer k-mesh of (551) was used for the density of state calculation of the $MoTe_2$(9×9)/$MoS_2$(10×10). The energy relaxation is continued until the changes of the total energy of the supercell and forces on all the atoms are converged to less than $10^{-4}$ eV/cell and $3\times10^{-2}$ eV Å$^{-1}$ respectively. Due to the extremely heavy weight of computation for such a large supercell, spin-orbit coupling was not taken into account. And it is for this reason, we choose to study a $MoTe_2$/$MoS_2$ bilayer, rather than a $WTe_2$/$MoS_2$ bilayer which has a very similar lattice mismatch but a much stronger spin-orbit coupling in the $WTe_2$ layer [38]. We have generated the preprocessing initial atomic structure and the post-processing band structure of supercells using the VASPKIT code [39].

### III. NUMERICAL RESULTS AND DISCUSSION

*A. Atomic layer corrugation*

**Fig. 1. Top (a) and side (b) views of the atomic structure of a $MoTe_2$/$MoS_2$ bilayer with (9×9)/ (10×10) stacking. *AA*, *AB*, and *BA* are the three local atomic configurations with three-fold rotational symmetry.**

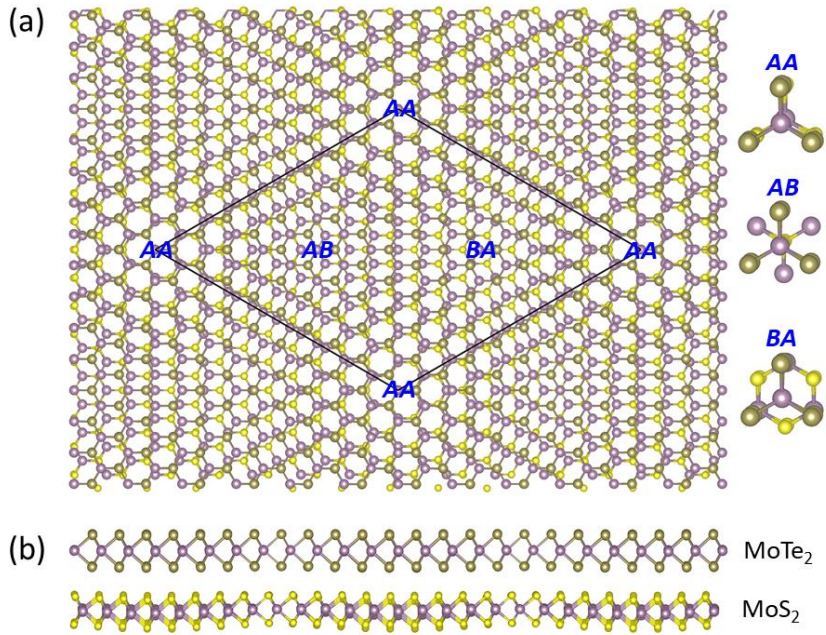

The top and side views of the atomic structure of a MoTe$_2$/MoS$_2$ bilayer with (9×9)/(10×10) stacking is shown in Fig. 1. Insets in panel (a) are close-ups of the three local atomic configurations with three-fold rotational symmetry, denoted as *AA*, *AB*, and *BA*. The configurations of (7×7)/(8×8), (8×8)/(9×9), and (10×10)/(11×11) stackings are shown in Fig. S1. Since MoTe$_2$ and MoS$_2$ monolayers have noticeably different lattice constants, a Moiré pattern appears (Fig. 1a) even when they are stacked one on top of the other, i.e., in the absence of twisting. Lattice corrugation, on the other hand, is also noticeable in the side view (Fig. 1b). The binding energy of each bilayer, defined as the energy lowering when pile two free-standing monolayers to form it, is displayed in Fig.2. as a function of superlattice misfit. Interestingly, we find that the data points can be fitted by a quadratic function fairly well, and the binding strength maximum deviates only slightly from the *nominal* zero-misfit point. The optimized lattice constants for MoTe$_2$ layer is 0.28% smaller than its free-standing monolayer, while that for MoS$_2$ is 0.09% larger than its free-standing counterpart. Such a pair of unbalanced changes is partially responsible for the disparity in corrugation amplitude of MoTe$_2$ and MoS$_2$ layers.

**Fig. 2. Interlayer binding strength of a MoTe$_2$/MoS$_2$ heterobilayer in different stacking as a function of the mismatch in superlattice. The red curve is a fitted quadratic function.**

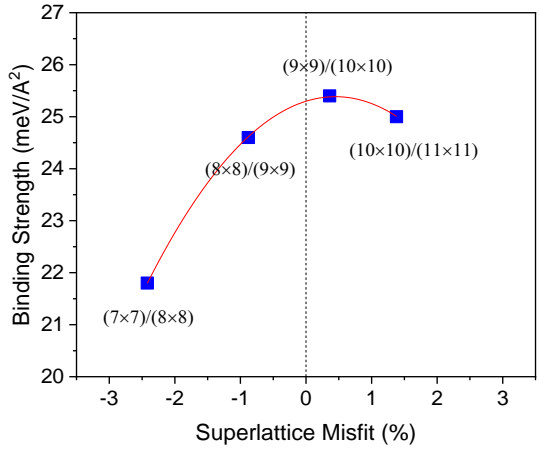

We display in Fig. 3 the height of Mo atoms (triangles) in each TMDC layer and also the interlayer distance between Mo atoms (squares), along the long diagonal of the superlattice. Dashed lines denote the high-symmetry locals in the Moiré cell. Note that only at *AA*, Mo atoms in MoTe$_2$ and MoS$_2$ are in exact on-top positions, thus their distance is precisely the interlayer distance. At other points, interpolation was used to obtain the interlayer distance. Figure 3 demonstrates clearly that both MoTe$_2$ and MoS$_2$ layers corrugates remarkably in a twist-free (9×9)/(10×10) stacking, with the latter having a larger amplitude (0.77Å) than the former (0.30Å). This is initially quite surprising to us because MoS$_2$ is supposed to be stretched and MoTe$_2$ to be compressed, as the (10×10) of MoS$_2$ supercell of has a smaller lattice than (9×9) of MoTe$_2$. Obviously, the interlayer vdW interaction exerts dissimilar effect on each monolayer, enlarging the lattice of MoS$_2$ and reducing that of MoTe$_2$. The stronger corrugation of MoS$_2$ accommodates a part of its enlarged 2D lattice constants, with respect to the MoTe$_2$ layer. The interlayer distance has the largest value in *AA* ($d_{AA}$=7.09Å), where Te and S atoms are on-top of each other; while at *AB* and *BA*, Te and S atoms are in each other's hollow site. $d_{AB}$ (6.59 Å) is a bit smaller than $d_{BA}$ (6.64 Å), because at *AB*, Mo in MoTe$_2$ is on top of S (smaller than Te in atomic size) in MoS$_2$, whereas at *BA*, Mo in MoS$_2$ is on top of Te in MoTe$_2$. The difference in $d_{AA}$ and $d_{AB}$ is 0.50 Å, a bit smaller than that obtained in [32] where local-density approximation (LDA) for the exchange-correlation interactions was employed. It has to be noted that the corrugation of MoS$_2$, although remarkable, is only about 2% of the supercell size and the related strain (change of the lattice constant) is as small as 0.08% (0.0025 Å).

It is worth noting that the calculated interlayer distance in the (1×1)/(1×1) stacking in *AA*, *AB* , and *BA* configurations is 7.20, 6.56, and 6.63 Å respectively. Interestingly, we find that $d_{AB}$ and $d_{BA}$ are very similar to the values in the (9×9)/(10×10) stacking, but the difference in $d_{AA}$ is remarkable. That is, the (1×1)/(1×1) stacking model overestimates the interlayer distance at the *AA* region by 0.11 Å.

**Fig. 3. The height of Mo atoms in MoTe$_2$ (upright triangles) and MoS$_2$ (upside down triangles) layers, and the interlayer distance (squares) along the diagonal in the Moiré cell. Data points are connected by spline lines to guide the eye.**

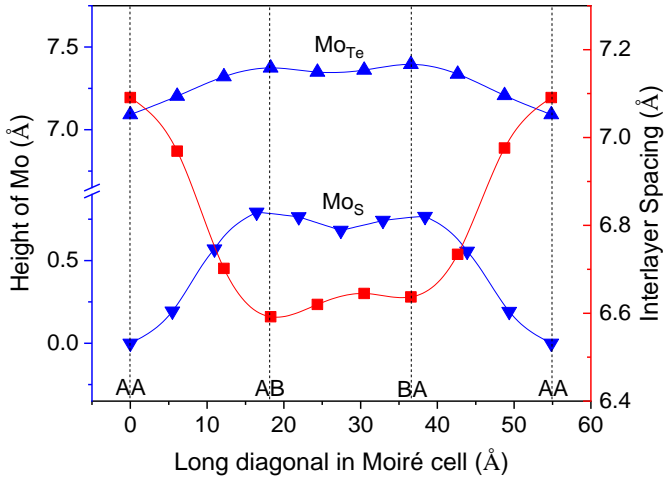

## *B. Strength of van der Waals interaction*

We note that the binding strength of a MoTe$_2$(9×9)/MoS$_2$(10×10) heterobilayer, 25.4 *m*eV/ Å$^2$, is significantly larger than that for the graphene bilayer, which is 11.2 *m*eV/ Å$^2$ for AB and 4.0 *m*eV/ Å$^2$ for *AA* stacking [40]. The stronger vdW interaction, together with significant lattice mismatch, renders a noticeable strain in each single layer in the MoTe$_2$(9×9)/MoS$_2$(10×10) heterostructure. On the other hand, such a strength is a bit smaller than that in the MoS$_2$ and MoTe$_2$ homobilayer, which is 27.2 and 28.4 meV/ Å$^2$ respectively, according to our calculation. This means that the effect of lattice mismatch on the interlayer vdW interactions, is around 10%. To evaluate the effect of corrugation on the interlayer binding, we have performed calculations on three restrained (hypothetical) configurations in which both MoTe$_2$ and MoS$_2$ layers were kept flat and only in-plane freedoms were relaxed, with the interlayer distance being the average $\bar{d}$, or $d_{AA}$, $d_{AB}$, respectively ($\bar{d}$=6.77 Å, $d_{AA}$=7.09 Å, $d_{AB}$=6.59 Å). The calculated binding strength of such a flat heterobilayer is 23.9 *m*eV/ Å$^2$ for $\bar{d}$, 23.06 *m*eV/ Å$^2$ for $d_{AA}$, and 21.2 *m*eV/ Å$^2$ for $d_{AB}$. Assuming a parabolic relationship between interlayer binding and spacing, we can estimate the maximum strength of a flat bilayer to be about 24.2 *m*eV/ Å$^2$ at $d = 6.88$Å. This is only 4.8% percent smaller than the fully relaxed, wrinkled configuration, indicating that corrugation has a noticeable but not significant effect on the interlayer binding.

**Fig. 4. Charge redistribution in the (110) plane of the Moiré cell, upon formation of a MoTe$_2$(9×9)/MoS$_2$(10×10) heterobilayer from monolayer MoTe$_2$ and MoS$_2$.**

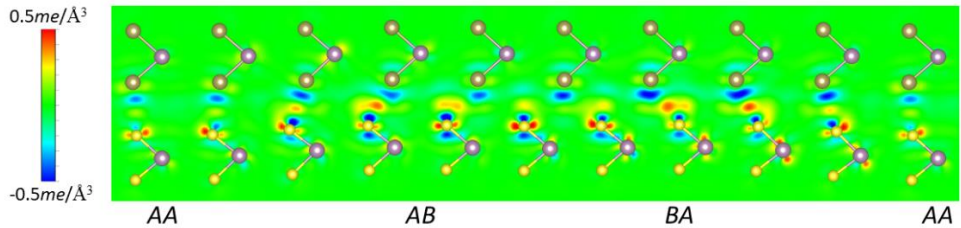

The charge redistribution associated with formation of a MoTe$_2$(9×9)/MoS$_2$(10×10) heterobilayer from monolayer MoTe$_2$ and MoS$_2$ is displayed in Fig. 4. From the amplitude of charge accumulation (red) and depletion (blue), the vdW bonding strength is clearly seen to increase form *AA* to *AB* and from *AA* to *BA*, due to a stronger repulsion in *AA* than in *AB* and *BA*. From *AB* to *BA*, nevertheless, the vdW bonding strength does not experience a monotonous change, but becomes weaker in between. This is because the Te-S distance around *AB* and *BA* is smaller than in between.

*C. Moiré pattern in local potential*

The French word *Moiré* refers to *an irregular wavy finish on a fabric*, which is essentially *an independent usually shimmering pattern seen when two geometrically regular patterns are superimposed especially at an acute angle* (Meriam-Webster). The Moiré pattern in optics was first investigated in 1874 by Lord Raleigh, who made use of the interference fringes resulted from two superimposed identical gratings to determine their quality even though each individual grating was too fine to be resolved under a microscope [41]. We note that the Moiré phenomenon is based solely on geometrical principles and independent of the image processing in the eye of the observer. In this sense, a moiré pattern is fundamentally the manifestation of a superstructure, either commensurate or incommensurate. In materials science, the Moiré superstructures scales down to superlattices. The physics involved in Moiré phenomena changes from simply the interference in an optical component to much more sophisticated chemical bonding or vdW interactions in a material system.

A Moiré pattern is visible when we look at the superimposed atomic structure of a MoTe$_2$(9×9)/MoS$_2$(10×10) heterobilayer shown in Fig. 1(a). The spatial variation of the 2D density of atomic spheres is noticeable, though not very conspicuous, in part due to the moderate size of the supercell. Switching from atomic structure to potential, we find that in the midplane of the interlayer, where the local potential has roughly the same contribution from constituent monolayers, the Moiré effect manifest itself prominently, as seen in Fig. 5. Strong contrast of light and dark regions reveals the signature of each individual monolayer. The AB and BA regions, for instance, are more distinguishable with opposite light-dark patterns than with 2D atomic density in the atomic structure (Fig. 1a). Unfortunately, the Moiré pattern of the local potential is hard to quantize in this plane because the two individual monolayers contribute equally here and none of their original unit cells can serve as the primary sublattice unit cell.

**Fig. 5. The midplane interlayer local potential in a MoTe$_2$/MoS$_2$ bilayer.**

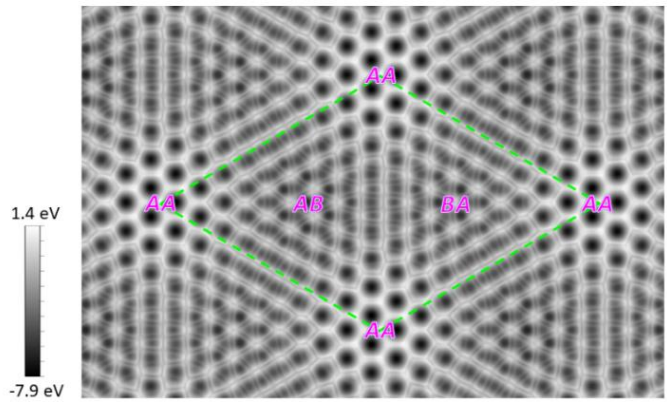

*D. Spatial variation of band structure*

Before discussing the band structure of the MoTe$_2$(9×9)/MoS$_2$(10×10) heterobilayer, we want to mention that our calculations demonstrate the (1×1)/(1×1) heterobilayer in *AA*, *AB*, and *BA* configurations are all gapless. It indicates that the large strain dramatically changed the electronic structure of the bilayer and therefore such a stacking model is unable to provide meaningful information even for the purpose of qualitative understanding of the MoTe$_2$/MoS$_2$ heterobilayer.

**Fig. 6. Species-projected band structure of a MoS$_2$/MoTe$_2$ bilayer with (a) and without (b) out-of-layer relaxation. Mo$_S$ and Mo$_{Te}$ represent Mo atoms in the MoS$_2$, MoTe$_2$ layers, respectively.**

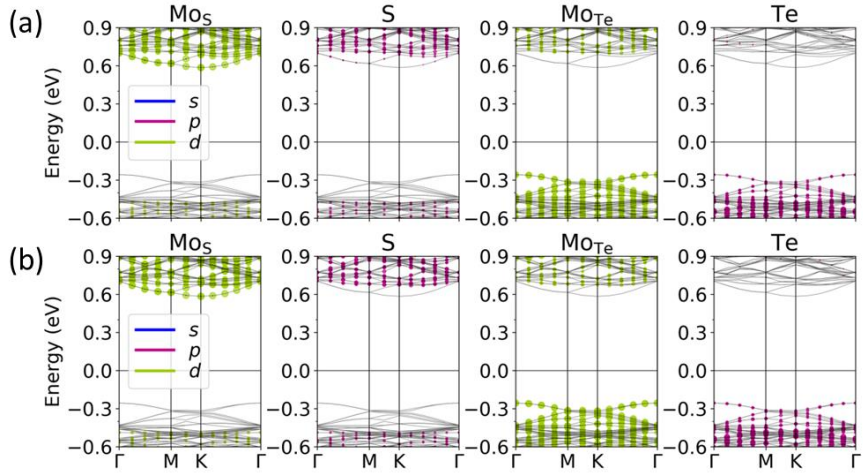

The calculated band structures of a twist-free MoTe$_2$(9×9)/MoS$_2$(10×10) heterobilayer near the Fermi level (set to zero), with (panel a) and without out-of-plane relaxation (pane b) are plotted in Fig. 6. Projections are made on *s*, *p*, *d* orbitals and also on each species. It is seen that corrugation influences the band structure only marginally, due to the weakness of vdW interactions. The valence band maximum (VBM) shifts down from -0.256 to -0.258eV, the conduction band minimum (CBM) shifts up from 0.584 to 0.585eV, and as a result the band gap increases slightly from 0.840 to 0.843eV (the LDA result is 0.78 eV [32]). Similar to the case of twisted MoTe$_2$($\sqrt{13} \times \sqrt{13}$)/MoS$_2$($\sqrt{16} \times \sqrt{16}$) bilayer which has a much smaller supercell [14], the twist-free MoTe$_2$(9×9)/MoS$_2$(10×10) bilayer is also a type II heterojunction with an indirect band gap. The valence band is mainly contributed by 4*d* orbitals of Mo in the MoTe$_2$ layer (denoted by Mo$_{Te}$) and partially by 5*p* states of Te. The conduction band consists of mainly 4*d* of Mo in MoS$_2$ layer (denoted by Mo$_S$) and the contribution from 3*p* states of S is insignificant. The VBM appears at the $\Gamma$ point, which corresponds to the K point in the unfolded Brillouin zone of a free-standing MoTe$_2$ monolayer, and the CBM is located at the K point, corresponding also to the K point of a free-standing MoS$_2$ monolayer. This means that the vdW interaction and the small artificial strain imposed in the supercell (-0.28% for MoTe$_2$ and +0.09% for MoS$_2$) have not changed the feature of direct band gap for both individual monolayers [31], unlike the large strain cases in MoS$_2$/WSe$_2$ bilayer [42].

**Fig. 7. Partial charge density isosurface (1.5×10$^{-4}$ *e*/Å$^3$) around the Mo atoms in the conduction (a) and valence (b) bands.**

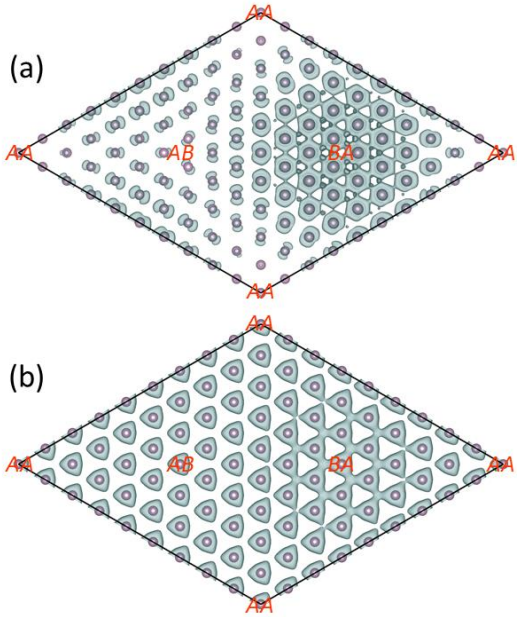

To study the spatial variation of the band gap of a MoTe$_2$(9×9)/MoS$_2$(10×10) bilayer, we here focus on the Moiré effect on the electronic states in the valence band of MoTe$_2$ and conduction band of MoS$_2$, both of which lie inside the gap of the other monolayer due to the band offsets and thus only weakly couple to states in that monolayer. The spatial variation of the valence and conduction bands can be well indicated by the distribution of electrons in those states. Figure 7 displays the partial charge density around Mo atoms in the conduction (panel a) and valence (panel b) bands. It is found that in both bands, the *BA* region has a higher charge density than *AA* and *AB* sites, especially in the former case, indicating that the Moiré pattern influence more significantly the electronic states in the MoS$_2$ layer. This is in accordance with the computational result that corrugation in the MoS$_2$ layer is more remarkable than in the MoTe$_2$ layer. A higher charge density in conduction band means a lower CBM for the Mo on that site, and a higher charge density in valence band suggests a higher VBM on that site. Therefore, we expect a smaller-than-average bandgap in the *BA* area.

**Fig. 8. Density of states of a MoTe$_2$/MoS$_2$ bilayer. The local DOS projected on the Mo atoms at or near *AA*, *AB*, and *BA* sites with three-fold rotational symmetry are drawn in bold lines.**

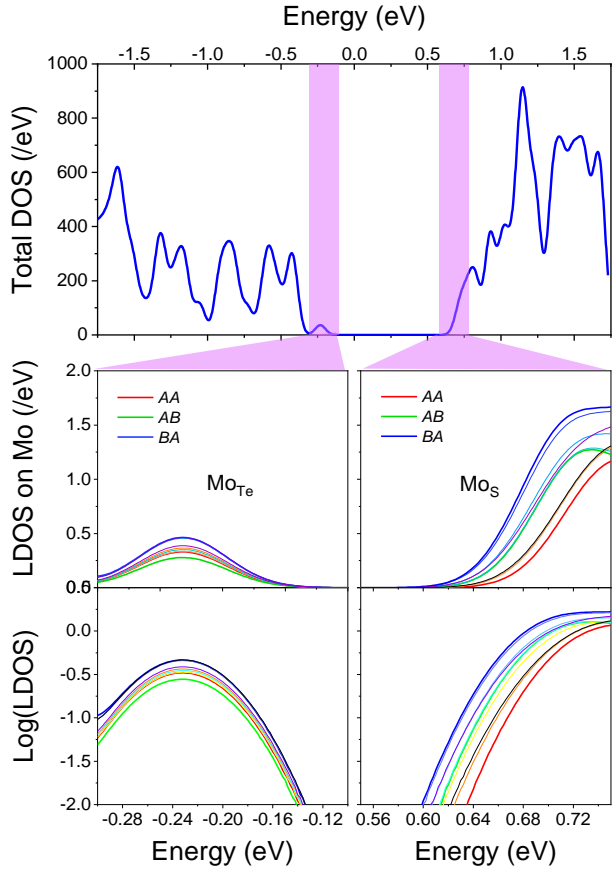

Although for each single band, we can figure out the contribution of each atom in the supercell, it is not straightforward to compose the local band structure in momentum space from such information for each atom. Therefore, we turn to the density of states (DOS) localized in a particular atom to obtain the its distribution in energy space, which fulfills our need for knowledge of band gap variation. In Fig. 8 we plot the total DOS of a $MoTe_2$(9×9)/$MoS_2$(10×10) bilayer (top) and the local DOS (middle) on the Mo atoms in narrow energy windows in both $MoTe_2$ ($Mo_{Te}$) and $MoS_2$ ($Mo_S$) monolayers along the diagonal in the Moiré cell (see Fig. 1a). Note that data for Mo located at AA, AB, and BA are plotted in bold curves to add emphasis. To make the displacement of LDOS curves more visible, we follow Carr et al. [6] to make use of the logarithm of LDOS (bottom panel) to estimate VBM and CBM for a selected number of Mo atoms. The DOS results of the flat bilayer, i.e., the configuration without out-of-plane relaxation are obtained in the same way (Fig. S2). In Fig. 9, we show the variation of the VBM (squares) and CBM (circles) in the Moiré cell with (solid symbols) and without (empty symbols) out-of-layer geometry relaxation. Data points are connected by spline lines to guide the eye. One of the salient features of the data is that the VBM varies much less strongly within the Moiré cell than does VBM. This is in accordance with the fact that $MoTe_2$ layer has dielectric constants about four times as large as that of $MoS_2$ [43], thus it is less susceptible to the electric field produced by Moiré pattern. Another finding is that the change of VBM has an opposite sign to that of CBM. We will see below that this is because the Moiré potential has opposite sign in these two monolayers. In addition, we find that the variation of CBM and VBM differs noticeably, but not significantly, in the wrinkled and flat bilayer. With the knowledge of VBM and CBM for different Mo atoms, we can readily obtain the spatial variation of band gap in the Moiré cell.

**Fig. 9. Variation of the valence band maximum (squares) and conduction band minimum (circles) in the Moiré cell with (solid symbols) and without (empty symbols) out-of-layer geometry relaxation. The average CBM and VBM is set to zero. Data points are connected by spine lines to guide the eye.**

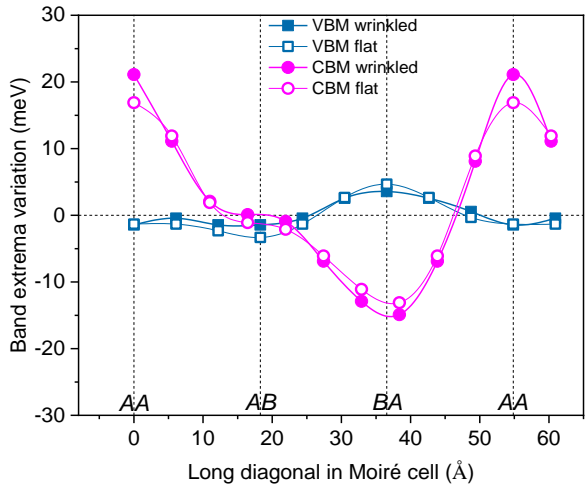

*E. Moiré potential*

As discussed above, since a bilayer vdW heterostructure is not strictly a 2D system, but rather composed of six atomic layers in the vertical direction, the definition of Moiré potential in real space, which is essentially a 2D function, depends on the surface or curved surface on which the phenomenon under investigation is present. Having realized that the valence and conduction bands are contributed mainly by $Mo_{Te}$ and $Mo_S$ atomic layer, respectively, and also inspired by the finding that total local potential in the interlayer midplane demonstrate well the Moiré pattern, we now examine the variation of the total local potential at the $Mo_{Te}$ and $Mo_S$ atoms. We argue that it is neither feasible nor necessary to trace the change of local potential at every single point in the primitive unit cell, a good representation could be the average of the local potential around the sole Mo atom therein. We plot in Fig. 10 the calculated average local potential at the $Mo_{Te}$ and $Mo_S$ atoms (atomic radius: 1.23Å) along the long diagonal of the Moiré cell, for both fully relaxed, wrinkled configuration (solid squares) and partially relaxed flat configuration (open squares). The data points are connected with spline lines to guide the eye, forming an envelope curve that indicates the amplitude of the local potential modulation. We can define the intralayer Moiré potential of the valence or conduction band as the variation of the corresponding average local potential in reference to its average over the Moiré cell.

Although the lattice corrugation is more significant in MoS$_2$ layer than in MoTe$_2$ layer (c.f. Fig. 3), its effect on Moiré potential is very similar for the two. In addition, Moiré pattern weakens upon geometry optimization, suggesting that it is part of the driving force for lattice relaxation. It has to be pointed out that lattice corrugation does not change the fundamental feature of the Moiré potential curve, but rather introduces a quantitative modification. Interestingly, as a function of the planar coordinates, the intralayer Moiré potential in Mo$_S$ always changes in opposite direction to that in Mo$_{Te}$. That is, when the Moiré potential goes up in the Mo$_S$ plane, it goes down in the Mo$_{Te}$ plane, and vice versa. The Moiré pattern in the average local potential in Mo$_S$ and Mo$_{Te}$ atomic planes are results of mutual modulation. Such an observation is in accordance with the discovery that the VBM and CBM (Fig. 9) also vary in anti-phase. The interlayer Moiré potential (or, simply termed as the Moiré potential) which relates to the band gap, can therefore be defined as the difference between the two intralayer Moiré potentials. Obviously, interlayer excitons - the Coulomb-bound states between electrons and holes located in different monolayers, are subject to the modulation of such a periodic potential.

**Fig. 10. The average local potential at the Mo$_{Te}$ and Mo$_S$ atoms. Solid squares display the data for fully relaxed, wrinkled configuration and empty squares represent partially relaxed, flat configuration. Data points are connected by spline lines to guide the eye.**

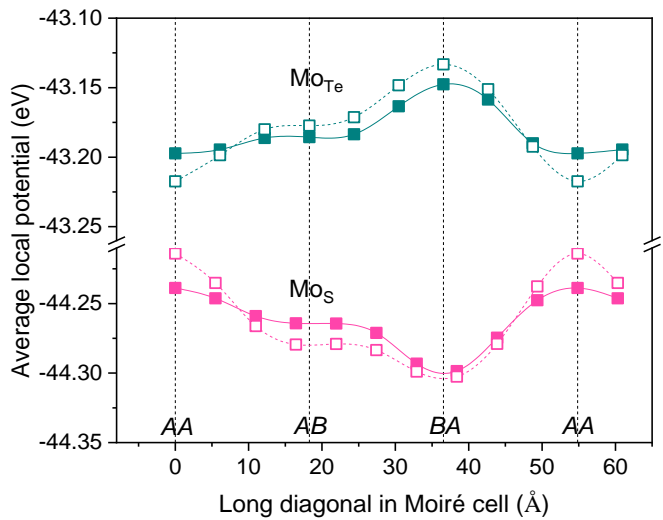

## F. Correlation of Moiré potential and band gap variation

For a more quantitative presentation, we plot the Moiré potential in Fig. 11 for both wrinkled (solid squares) and flat (empty squares) MoTe$_2$/MoS$_2$ bilayers. It is found that corrugation reduces remarkably the amplitude of the Moiré potential, from 0.17 to 0.11 eV. We also display in Fig. 11 the change of band gap along the diagonal of the Moiré cell. It has an amplitude of 0.04 eV. In contrast to its effect on the Moiré potential, lattice corrugation of the bilayer enhances the spatial variation of the local band gap by 5 meV. On the other hand, its influence on the global band gap is within 1 meV.

**Fig. 11. Correlation of the Moiré potential (squares) and bandgap variation (circles) in a MoTe$_2$/MoS$_2$ bilayer with (solid symbols) and without (empty symbols) out-of-layer geometry relaxation. Data points are connected by spline lines to guide the eye.**

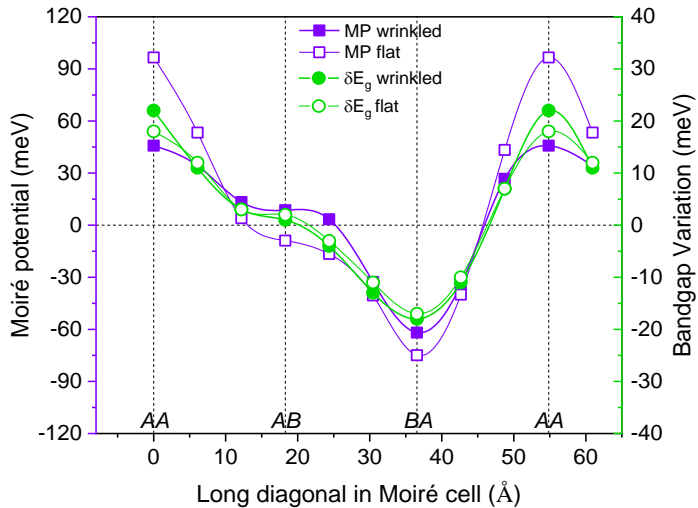

Clearly, there is a strong and direct correlation between the variation of Moiré potential and the band gap in the Moiré cell. They are roughly proportional to each other. This is an indication that the valence band of MoTe$_2$ couples only weakly to the MoS$_2$ layer, and likewise, the conduction band of MoS$_2$ interacts also weakly to the MoTe$_2$ layer. The deviation from a proportional relationship can be partially understood from the species-projected band plot in Fig. 5, in which we find the Te atoms have a noticeable contribution to the valence band. The Moiré potential defined for the Mo$_{Te}$ atomic layer is therefore not solely responsible for the spatial variation of the valence band Moiré potential, and the contribution of the Moiré pattern of the average local potential at the Te atoms should also be included in a more precise treatment. The depth of the Moiré well, or, Moiré trap, formed by spatial variation of Moiré potential, is dependent on the local details of the interlayer vdW interactions. And in this sense, the Moiré potential can be directly related the chemistry of vdW heterostructures.

To understand the correlation between the spatial variation of band gap and Moiré potential in the Morié cell, we propose a simple intralayer band-coupling model. The band gap of a weakly coupled vdW semiconducting heterobilayer depends mainly on the band edges of upper and lower layers as well as their alignments. As an approximation, here we consider only the states of conduction (indexed by $c$) and valence band ($v$) edges of the upper ($u$) and lower ($l$) layer, represented by $\psi^u = (\psi_c^u, \psi_v^u)$, $\psi^l = (\psi_c^l, \psi_v^l)$, respectively. Correspondingly, the states of the heterobilayer system can be described by wave function $\Psi = (\psi^u, \psi^l) = (\psi_c^u, \psi_v^u, \psi_c^l, \psi_v^l)$. The Hamiltonian consists of three parts, $\hat{H} = \hat{H}^u + \hat{H}^l + \hat{V}$, where $\hat{H}^u, \hat{H}^l$ denote the Hamiltonian of upper, lower layers, and $\hat{V}$ denotes the perturbative Hamiltonian caused by vdW interaction. The matrix of Hamiltonian in this representation reads

$$H = \langle \Psi | \hat{H} | \Psi \rangle$$

$$= \begin{pmatrix} \langle \psi_c^u | \hat{H}^u + \hat{V} | \psi_c^u \rangle & \langle \psi_c^u | \hat{H}^u + \hat{V} | \psi_v^u \rangle & \langle \psi_c^u | \hat{V} | \psi_c^l \rangle & \langle \psi_c^u | \hat{V} | \psi_v^l \rangle \\ \langle \psi_v^u | \hat{H}^u + \hat{V} | \psi_c^u \rangle & \langle \psi_v^u | \hat{H}^u + \hat{V} | \psi_v^u \rangle & \langle \psi_v^u | \hat{V} | \psi_c^l \rangle & \langle \psi_v^u | \hat{V} | \psi_v^l \rangle \\ \langle \psi_c^l | \hat{V} | \psi_c^u \rangle & \langle \psi_c^l | \hat{V} | \psi_v^u \rangle & \langle \psi_c^l | \hat{H}^l + \hat{V} | \psi_c^l \rangle & \langle \psi_c^l | \hat{H}^l + \hat{V} | \psi_v^l \rangle \\ \langle \psi_v^l | \hat{V} | \psi_c^u \rangle & \langle \psi_v^l | \hat{V} | \psi_v^u \rangle & \langle \psi_v^l | \hat{H}^l + \hat{V} | \psi_c^l \rangle & \langle \psi_v^l | \hat{H}^l + \hat{V} | \psi_v^l \rangle \end{pmatrix}. \quad (2)$$

Since $\psi_\alpha^i$ ($i = (u, l), \alpha = (c, v)$) are the eigen wavefunctions of $\hat{H}^i$, we have $\langle \psi_\alpha^i | \hat{H}^i | \psi_\beta^i \rangle \overset{\text{def}}{=} \varepsilon_\alpha^i \delta_{\alpha\beta}$ for orthogonality, where $\varepsilon_\alpha^i$ is the $\alpha$ band edge energy in the $i$th layer and $\delta_{\alpha\beta}$ is the Kronecker symbol. As seen in Fig. 6, the band edge states of both MoS$_2$ and MoTe$_2$ layers are mainly constituted by Mo-$d$ orbits. The Mo$_{Te}$ and Mo$_S$ atomic planes, or, sublayers, are in fact sandwiched by two chalcogens sublayers, thus, the overlap of their wavefunctions is negligibly small. Therefore, we set $\langle \psi_\alpha^u | \hat{V} | \psi_\beta^l \rangle = \langle \psi_\beta^l | \hat{V} | \psi_\alpha^u \rangle^* = 0$. On the other hand, the intralayer coupling depends on the symmetry of wavefunctions as well as the overall interlayer vdW coupling. For both MoS$_2$ and MoTe$_2$ single-layer, the states at the top of valence band involve only $d_{x^2-y^2}$ and $d_{xy}$ of Mo and those at the bottom of conduction bands contain mainly $d_{z^2}$ of Mo. The point group of the three high-symmetry local atomic configurations (*AA*, *AB*, *BA*) in MoS$_2$/MoTe$_2$ heterobilayer are all C$_{3v}$. The irreducible representation of the states at the top of valence bands and at the bottom of conduction bands are two-dimensional representation E and the totally symmetric representation A$_1$, respectively. The intralayer coupling induced by the interlayer vdW interaction is essentially the interaction between the induced dipoles of the d-electrons at band edges and the polarized electric field arising from the charge redistribution of chalcogens atoms (c.f. Fig. 4).

The intralayer perturbation Hamiltonian is of the form $\hat{V} = -\hat{\vec{P}} \cdot \vec{E} = -\vec{E} \cdot \hat{\vec{\alpha}} \cdot \vec{E}$, where $\vec{E}$ is the polarized electric field vector, $\hat{\vec{P}} = \hat{\vec{\alpha}} \cdot \vec{E}$ is the induced dipole moment operator of band edge $d$-electrons of Mo atoms, and $\hat{\vec{\alpha}}$ is the polarizability operator, a symmetric second tensor operator with six independent matrix elements. The symmetry of $\hat{V}$ is determined by $\hat{\vec{\alpha}}$. Under the symmetry of point group C$_{3v}$, its representation is reducible, and can be expressed by the direct sum of irreducible representations of C$_{3v}$ as A$_1$+A$_2$+2E. One can prove that this kind of vdW interaction would lead to both intra-band and inter-band coupling of conduction and valence bands in every layer. For simplicity, we denote the intralayer coupling matrix elements $\langle \psi_\alpha^i | \hat{V} | \psi_\alpha^i \rangle = -\vec{E}^i \cdot \hat{\vec{\alpha}}_\alpha^i \cdot \vec{E}^i = \mu_\alpha^i$, $\langle \psi_\alpha^i | \hat{V} | \psi_\beta^i \rangle = \langle \psi_\beta^i | \hat{V} | \psi_\alpha^i \rangle^* = \tau^i$, where $i = (u, l), \alpha = (c, v)$. In term of these parameters, the Hamiltonian matrix $H$ can be simplified to a more compact form,

$$H = \begin{pmatrix} \varepsilon_c^u + \mu_c^u & \tau^u & 0 & 0 \\ \tau^{u*} & \varepsilon_v^u + \mu_v^u & 0 & 0 \\ 0 & 0 & \varepsilon_c^l + \mu_c^l & \tau^l \\ 0 & 0 & \tau^{l*} & \varepsilon_v^l + \mu_v^l \end{pmatrix}. \quad (3)$$

The eigen equation can be resolved and the energy levels for upper and lower layers are

$$E_{c/v}^u = \frac{1}{2}(\varepsilon_c^u + \varepsilon_v^u + \mu_c^u + \mu_v^u) \pm \frac{1}{2}\sqrt{(\varepsilon_g^u + \mu_g^u)^2 + 4|\tau^u|^2} \quad (4)$$

and

$$E_{c/v}^l = \frac{1}{2}(\varepsilon_c^l + \varepsilon_v^l + \mu_c^l + \mu_v^l) \pm \frac{1}{2}\sqrt{(\varepsilon_g^l + \mu_g^l)^2 + 4|\tau^l|^2}, \quad (5)$$

where $\varepsilon_g^u = \varepsilon_c^u - \varepsilon_v^u$, $\mu_g^u = \mu_c^u - \mu_v^u$, $\varepsilon_g^l = \varepsilon_c^l - \varepsilon_v^l$, $\mu_g^l = \mu_c^l - \mu_v^l$. For the MoTe$_2$/MoS$_2$ bilayer, the CBM and VBM of MoTe$_2$ are sitting higher than those of MoS$_2$ respectively, in a type II band alignment. The bandgap is equal to the energy difference between the CBM of MoS$_2$ monolayer and the VBM of MoTe$_2$ monolayer,

$$E_g = E_c^l - E_v^u = \frac{1}{2}\left[(\varepsilon_c^l + \varepsilon_v^l + \mu_c^l + \delta_v^l) - (\varepsilon_c^u + \varepsilon_v^u + \mu_c^u + \mu_v^u)\right]$$

$$+ \frac{1}{2}\left(\sqrt{(\varepsilon_g^l + \mu_g^l)^2 + 4|\tau^l|^2} + \sqrt{(\varepsilon_g^u + \mu_g^u)^2 + 4|\tau^u|^2}\right). \quad (6)$$

Since $|\tau^i| \ll (\varepsilon_g^i + \mu_g^i), i = (u, l)$, the bandgap can be approximated as

$$E_g \approx \varepsilon_g + \mu_g + \frac{|\tau^l|^2}{(\varepsilon_g^l + \mu_g^l)} + \frac{|\tau^u|^2}{(\varepsilon_g^u + \mu_g^u)}, \quad (7)$$

where $\varepsilon_g = \varepsilon_c^l - \varepsilon_v^u$, $\mu_g = (\mu_c^l - \mu_v^u) = -(\vec{E}^l \cdot \widehat{\vec{\alpha}_c} \cdot \vec{E}^l - \vec{E}^u \cdot \widehat{\vec{\alpha}_v} \cdot \vec{E}^u)$. Since the $\mu_\alpha^i$ and $\tau^i$ are of same order and small compared with $\varepsilon_g^i$, the last two terms in the bandgap can be further neglected. Knowing that $\mu_\alpha^i$ are spatial function in the Mo atomic plane due to the presence of Moiré pattern, we denote their average as $\overline{\mu}_\alpha^i$, then $\mu_\alpha^i = \overline{\mu}_\alpha^i + \Delta\mu_\alpha^i$. The variation of band gap becomes

$$\Delta E_g = E_g - \bar{E}_g \approx \Delta\mu_\alpha^i = -2(\overline{\vec{E}^l} \cdot \widehat{\vec{\alpha}_c} \cdot \Delta\vec{E}^l - \overline{\vec{E}^u} \cdot \widehat{\vec{\alpha}_v} \cdot \Delta\vec{E}^u) \tag{8}$$

where $\overline{\vec{E}^l}$ is the average intensity of the electric field with an variation of $\Delta\vec{E}^l$.

Firstly, the *d*-electron polarization in MoS$_2$ layer is stronger than in MoTe$_2$ (c.f. Fig. 4), yielding a more significant energy variation in CBM than in VBM, as shown in Fig. 9. Thus the band gap variation is dominated by $\Delta\mu_c^l = -2\overline{\vec{E}^l} \cdot \widehat{\vec{\alpha}_c} \cdot \Delta\vec{E}^l$. Secondly, the intensity of the polarized electric field induced by polarization of S and Te atoms inside the bilayer varies with the local atomic structure. The electric field intensity is low in *AA* area, high in *BA*, and medium in *AB*. This means that $\Delta\vec{E}^l$ is antiparallel in *AA* and parallel in *BA* to $\overline{\vec{E}^l}$; while in *AB*, $\Delta\vec{E}^l$ is relatively small. As a result, the band gap has a maximum in *AA* and a minimum in *BA* area; while in *AB* area it is close to the average, as shown in Fig. 11. Such a rough estimation is in fairly good agreement with the comparison of the band gap and Moiré Potential shown in Fig. 11, which demonstrates that the bandgap variation is roughly proportional to the Moiré Potential.

**IV. CONCLUDING REMARKS**

In a transition metal dichalcogenide heterobilayer, with small lattice mismatch and/or twisting, the Moiré potential mixes momentum states separated by Moiré reciprocal lattice vectors and produces satellite optical absorption peaks, resulting the emergence of Moiré exciton [26]. To have a precise description of the phenomena related to Moiré pattern, a fully *ab initio* determination of the electronic structure of such a bilayer is indispensable. Our density functional theory calculations on a MoTe$_2$(9×9)/MoS$_2$(10×10) system, which has a moderate superlattice larger than an exciton, but not large enough to justify a continuum model, have allowed full lattice relaxation. We find that variation in the average of the local potential in the Mo layers can serve as an excellent illustration of the Moiré pattern in the van der Waals heterostructure. The interlayer real-space Moiré potential, defined as the difference in the variation of local potential average between the Mo atomic layers, has a depth of 0.11 eV and might be effective to trap interlayer excitons. In addition, it changes approximately in proportion to the band gap variation in the Moiré cell. Although lattice corrugation is remarkable in both MoTe$_2$ (0.30Å) and MoS$_2$ (0.77Å) layers, the curvature of the wrinkled monolayers is rather small due to the large size of the Moiré cell. Therefore, it increases the spatial variation of the local band gap only by 5 meV, and its influence on the global band gap is no larger than 1 meV.

Our DFT calculations also demonstrate that the (1×1)/(1×1) stacking model of this heterobilayer yields a metallic electronic structure. It strongly signifies the necessity of employing a realistic supercell to minimize the lattice match, in order to have a precise description of the bilayer. Since the interlayer coupling can be modified by external vertical electric field [44] and by pressure [45], the Moiré potential is experimentally tunable. The Moiré potential defined in the present work differs from the one introduced by Wu et al. [26] in that our potential can be obtained before the determination of the spatially varying band gap, and can be more readily correlated to chemical bonding in individual monolayer and vdW interactions across the bilayer. As a consequence, it might be more useful in the practice of chemical engineering of two-dimensional vdW structures.

**ACKNOWLEDGMENTS**

This work was supported by Innovative Science and Technology Initiative for Security, ATLA, Japan.

**Graphical Abstract**

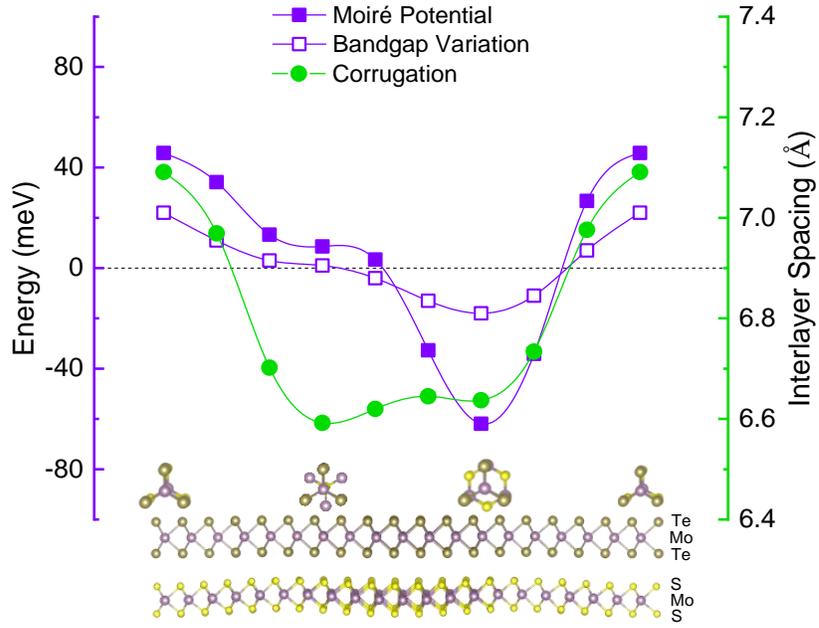

# Supplementary Materials

# Moiré potential, lattice corrugation, and band gap variation in a twist-free MoS$_2$/MoTe$_2$ heterobilayer

W. T. Geng[a‡], V. Wang[b], T. Ohno[a], J. Nara[a§]

[a] *National Institute for Materials Science, Tsukuba 305-0044, Japan.*
[b] *Department of Applied Physics, Xi'an University of Technology, Xi'an 710054, China*

**Fig. S1.** Top and side views of the atomic structure of a MoTe$_2$/MoS$_2$ bilayer with (7×7)/(8×8) [panel (a)], (8×8)/(9×9) [panel (b)], and (10×10)/(11×11) [panel (c)] stackings. Lattice corrugation in the MoS$_2$ layer is clearly seen.

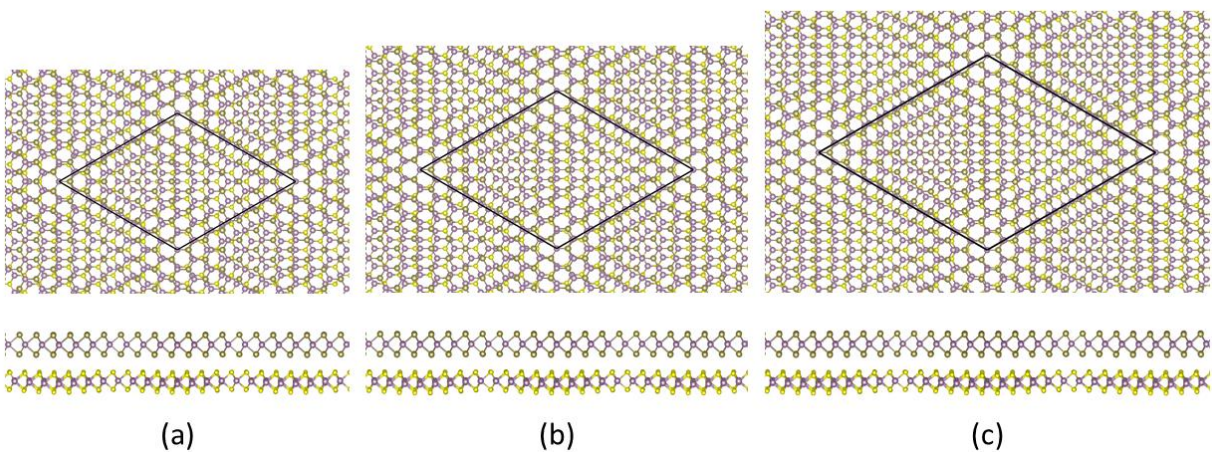

(a)  (b)  (c)

---

‡ geng.wentong@nims.go.jp
§ nara.jun@nims.go.jp

**Fig. S2. Density of states of a MoTe$_2$/MoS$_2$ bilayer.** The local DOS projected on the Mo atoms at or near *AA*, *AB*, and *BA* sites with three-fold rotational symmetry are drawn in bold lines.

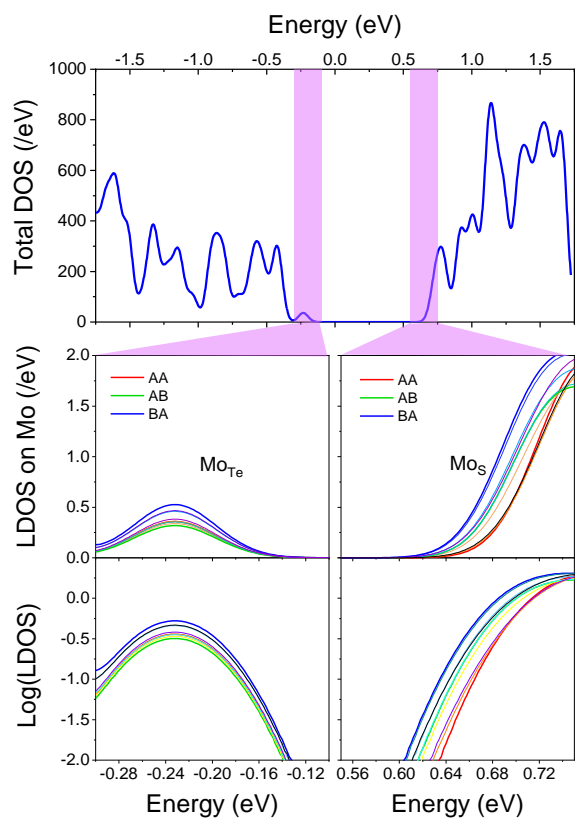